\documentclass[twocolumn,preprintnumbers,amssymb,amsmath,superscriptaddress,letterpaper]{revtex4}[12pt]
\usepackage{times,graphics}
\usepackage[colorlinks]{hyperref}

\usepackage{graphicx}
\usepackage{dcolumn}
\usepackage{bm}
\usepackage{natbib}
\usepackage{pstricks}
\usepackage{epsfig}
\usepackage{epstopdf}
\usepackage{multirow}

\newcommand{\be}{\begin{equation}}
\newcommand{\ee}{\end{equation}}
\newcommand{\bea}{\begin{eqnarray}}
\newcommand{\eea}{\end{eqnarray}}

\begin{document}

%\title{ $\Lambda\oplus$CDM\\ $H_0$ tension as a hint for \"Uber-Gravity}
%\title{$H_0$ tension as a hint for \"Uber-Gravity}
\title{$H_0$ tension as a hint for a transition in gravitational theory}
\author{Nima Khosravi}
\email{n-khosravi@sbu.ac.ir}
\affiliation{Department of Physics, Shahid Beheshti University, G.C., Evin, Tehran 19839, Iran}
\author{Shant Baghram}
\email{baghram@sharif.edu}
\affiliation{Department of Physics, Sharif University of Technology, P. O. Box 11155-9161, Tehran, Iran}
\author{Niayesh Afshordi}
\email{nafshordi@pitp.ca}
\affiliation{Perimeter Institute for Theoretical Physics, 31 Caroline St. N., Waterloo, ON, N2L 2Y5, Canada}
\affiliation{Department of Physics and Astronomy, University of Waterloo, Waterloo, ON, N2L 3G1, Canada}
\author{Natacha Altamirano}
\email{naltamirano@pitp.ca}
\affiliation{Perimeter Institute for Theoretical Physics, 31 Caroline St. N., Waterloo, ON, N2L 2Y5, Canada}
\affiliation{Department of Physics and Astronomy, University of Waterloo, Waterloo, ON, N2L 3G1, Canada}

\date{\today}

\begin{abstract}

%It is a theory! It is not a model!
We propose a cosmological model, \"u$\Lambda$CDM, based on {\it \"uber-gravity}, which is a canonical ensemble average of many theories of gravity. In this model, we have a sharp transition from (a purely) $\Lambda$CDM era to a phase in which the Ricci scalar is a constant. This transition occurs when the Ricci scalar reaches a critical scale or alternatively at a transition redshift, $z_\oplus$. We use the observations of baryonic acoustic oscillations (BAO) and Supernovae Ia (SNe), as well as the cosmic microwave background (CMB) data to constrain \"u$\Lambda$CDM. This yields $H_0=70.6_{-1.3}^{+1.1}$ km/s/Mpc, $\Omega_m=0.2861 \pm{0.0092} $ and $z_\oplus=0.537_{-0.375}^{+0.277}$, providing a marginally better fit with a Akaike information criterion of 0.8. %After adding in the data from BAO in the Lyman-$\alpha$ forest in quasar spectra, we get $\Delta\chi^2=-2.14$, where the best fit occurs at $H_0=71.2_{-1.5}^{+1.8}$ km/s/Mpc, $\Omega_m=0.280\pm 0.013$ and $z_{\oplus}= 0.41^{+0.31}_{-0.29}$.
Therefore, \"u$\Lambda$CDM can ease the $H_0$-tension, albeit marginally, with one additional free parameter. We also provide a preliminary study of the linear perturbation theory in \"u$\Lambda$CDM which points to interesting potential {\it smoking guns} in the observations of large scales structure at $z < z_{\oplus}$.
\end{abstract}
\maketitle
\section{Introduction:}
The standard model of cosmology, $\Lambda$CDM, consists of well-known baryons, unknown cold dark matter (CDM) and dark energy, which is represented by the cosmological constant ($\Lambda$). Also the gravity is governed by the Einstein general relativity (GR) in the standard model. The vanilla $\Lambda$CDM model is a favored one as it fits well almost all of the observations such as cosmic microwave background radiation (CMB) \cite{Ade:2015xua} and large scale structure (LSS) \cite{Tegmark:2003ud}. However, fundamental questions remain, e.g., what are the natures of dark energy and dark matter? What is responsible for a highly fine-tuned cosmological constant, in comparison to the vacuum energy density predicted by the particle physics (otherwise known as the cosmological constant problem)?

On the observational side, notable tensions between best fit $\Lambda$CDM theoretical predictions \cite{Ade:2015xua} and data remain, which include: $H_0$ tension \cite{Riess:2016jrr,riess18,Riess:2019cxk,Bernal:2016gxb}, $\sigma_8$ tension \cite{Abbott:2017wau,Joudaki:2017zdt,Joudaki:2016mvz}, BAO in the Lyman-$\alpha$ forest \cite{Bourboux:2017cbm}, void phenomenon \cite{Peebles:2001nv} and missing satellite problem \cite{Klypin:1999uc}. While such tensions can be (and often are) due to systematic errors, some may provide clues to the physics beyond the standard models of cosmology and particle physics. As an interesting idea which has been studied to address both $H_0$ and $\sigma_8$ tensions is massive neutrinos \cite{Poulin:2018zxs}. To lessen $H_0$ tension different approaches have been extensively studied in the literature of modified gravity (e.g., \cite{Amendola:2016saw,Copeland:2006wr,Jain:2010ka,Clifton:2011jh}) including interacting dark energy \cite{DiValentino:2017iww,Yang:2018euj}, neutrino-dark matter interaction \cite{DiValentino:2017oaw},  varying Newton constant \cite{Nesseris:2017vor}, viscous bulk cosmology \cite{Mostaghel:2016lcd}, phantom-like dark energy \cite{DiValentino:2016hlg}, early dark energy \cite{Poulin:2018cxd}, massive graviton \cite{DeFelice:2016ufg}, phase transition in dark energy \cite{Banihashemi:2018has,Banihashemi:2018oxo}, decaying dark matter \cite{Vattis:2019efj}, etc. As another example, warm dark matter as an idea with some roots in particle physics has been proposed as  a solution for missing satellite problem \cite{Bode:2000gq}. However, none of the above alternatives has been quite as compelling as $\Lambda$CDM.

Here we pursue a different perspective on this problem: In spite of (presumable) existence of a huge number of distinct theoretically consistent models, how can Nature only pick one? This leads to an idea described in \cite{Khosravi:2016kfb} and based on that a model, \"uber-gravity, introduced in \cite{Khosravi:2017aqq} which we will briefly review in the following section. We then study the resulting cosmology and show that it has a rich phenomenology, with the potential to address the $H_0$ and BAO in the Lyman-$\alpha$ forest tensions, as well as distinct predictions for structure formation at low redshifts.
The structure of this work is as below: In Sec.(\ref{Sec:2}), we introduce the idea of ensemble average theory of gravity and the corresponding  \"uber-gravity model. In Sec.(\ref{Sec:3}), we introduce the cosmological model that follows \"uber-gravity, which we call \"u$\Lambda$CDM.   In Secs.(\ref{Sec:b}) and (\ref{pert}), we study the background and perturbation of \"u$\Lambda$CDM. Finally in Sec.(\ref{Sec:4}), we conclude and remark on future directions.
% ********************************************
\section{\"Uber-Gravity}
\label{Sec:2}
In this section, we review the idea of the ensemble average theory of gravity and \"uber-gravity in the upcoming two subsections respectively.
\subsection{Ensemble Average Theory of Gravity}
The ``Ensemble Average Theory of Gravity" \cite{Khosravi:2016kfb} suggests that the gravity model is the average over all the theoretically possible models of gravity. For this reason, a recipe has been suggested which is inspired by path integral formalism. This idea has some relationship with the ``Mathematical Universe" idea of Tegmark \cite{Tegmark:2007ud}. The same philosophy has been used in the context of particle physics by Arkani-Hamed et al. \cite{Arkani-Hamed:2016rle}. In \cite{Arkani-Hamed:2016rle}, it is assumed that there are different types of standard model of particle physics labeled by their Higgs masses. The idea of taking averaging over all the possible models can give a clue to address hierarchy problems \cite{Khosravi:2017aqq,Arkani-Hamed:2016rle}. % and \cite{Khosravi:2017aqq}.
To implement this idea we suggest to work with a Lagrangian which has been defined as \cite{Khosravi:2016kfb}
\begin{eqnarray} \label{EAT-lagrangian}
{\cal L}= \bigg(\displaystyle\sum_{i=1}^{N}{\cal L}_i e^{- \beta{\cal L}_i}\bigg) \bigg/ \bigg(\displaystyle\sum_{i=1}^{N} e^{- \beta{\cal L}_i}\bigg),
\end{eqnarray}
where ${\cal L}_i$'s are the theoretically possible Lagrangians and $\beta$ is a free parameter of this model. As is commonly done in statistical physics, we can write this as:
\begin{eqnarray}\label{partition}
{\cal L}=-\frac{\partial }{\partial \beta}\,\ln{\cal Z}, \hspace{2cm}{\cal Z}\equiv\sum_{i=1}^{N}\, e^{-\beta {\cal L}_i}
\end{eqnarray}
where ${\cal Z}$ is the canonical partition function in the model space.
In the next subsection, we will use the idea to make a toy model.

\subsection{\"Uber-Gravity model}
\begin{figure}
	\centering
	\includegraphics[width=1\linewidth]{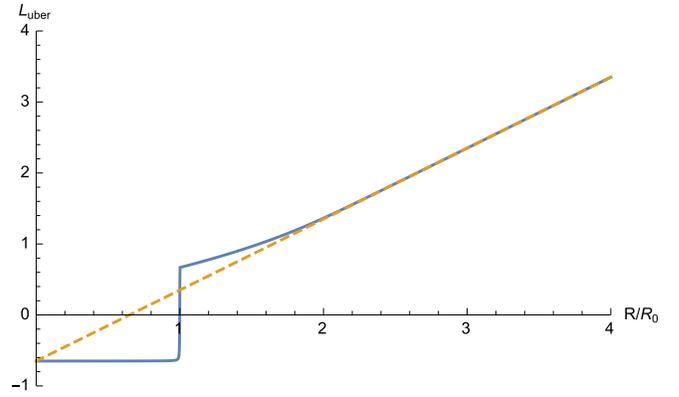}
	\caption{Blue line is our Lagrangian (\ref{ubergravity-Lagrangian}) for $\Lambda=0.32\,R_0$ and $\beta=2.5$ where we do sum up to $N=1000$ (It is easy to see that for larger $N$'s the plot is practically the same.) and yellow dashed line shows standard EH action with the same value for $\Lambda$.}
	\label{fig:fR1}
\end{figure}

In \cite{Khosravi:2017aqq}, the above idea has been used in the context of gravity and here we will review it very briefly. Let's define the partition function over the all analytic models of gravity as
\begin{eqnarray}\label{partition-fR}
{\cal Z}=\sum_{f(R)}\, e^{-\beta f(R)},
\end{eqnarray}
where $f(R)$'s are analytic functions of Ricci scalar, $R$. In \cite{Khosravi:2017aqq}, it has been shown that the final Lagrangian, dubbed \"uber-gravity, is not sensitive to the choice of basis for its main properties. In general, for analytic functions of $f(R)$ we can set basis as $\alpha_n R^n+ \lambda_n$ for each $n \in \mathbb N$. For simplicity, here we focus on $\alpha_n=1/R_0^{n}$ and $\lambda_n=-2\Lambda$, which yields:
\begin{eqnarray}\label{ubergravity-Lagrangian}
{\cal L}_{{\rm \ddot{u}ber}}= \bigg(\displaystyle\sum_{n=1}^{\infty}(\bar{R}^n-2\Lambda) e^{- \beta (\bar{R}^n-2\Lambda)}\bigg) \bigg/ \bigg(\displaystyle\sum_{n=1}^{\infty} e^{- \beta (\bar{R}^n-2\Lambda)}\bigg),
\end{eqnarray}
where ${\bar R}\equiv R/R_0$ and $R_0$ is a new free parameter of the model with dimension $[M^2]$ which makes $\beta$ dimensionless. An example of ${\cal L}_{{\rm \ddot{u}ber}}$ is shown in  Fig. \ref{fig:fR1}.

\begin{figure}
	\centering
	\includegraphics[width=1\linewidth]{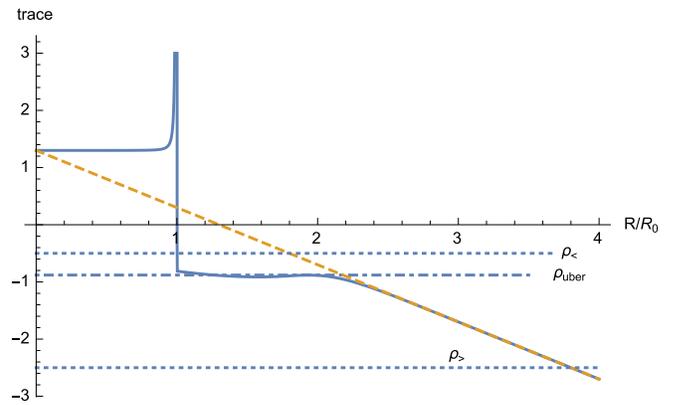}
	\caption{Blue line is the trace of equation of motion in \"uber-gravity where $\Lambda=0.32\,R_0$, $\beta=2.5$ and yellow dashed line shows the same for the EH action. For $\rho > \rho_{{\rm \ddot{u}ber}}$ the matter field sees gravity as standard EH action and for $\rho < \rho_{{\rm \ddot{u}ber}}$ the gravity switches to $R=R_0$.}
	\label{fig:trace-full}
\end{figure}

The \"uber-gravity has the following universal properties: i) for high-curvature regime it reduces to the Einstein-Hilbert (EH) action i.e. $R-2\Lambda$, ii) for intermediate-curvature regime it predicts a stronger gravity than the EH model, iii) it is vanishing  for low-curvature regime ($R<R_0$) and iv) there is a sharp transition at $R_0$,
which is not sensitive to choice of the basis and parameters \cite{Khosravi:2017aqq}. In this sense, \"uber-gravity is a fixed point in the model space which makes it unique. The main goal of this work is to study the cosmology of our model and for this purpose we need to study the equations of motion. However, for our purpose we need the trace of equation of motion (and we assume the case of steady state i.e. $R$ is evolving very slowly) which is plotted in Fig. \ref{fig:trace-full}.
In the next section, we will introduce a cosmological model based on the general behavior of the \"uber-gravity.

%********************************************
\section{\"u$\Lambda$CDM cosmology}
\label{Sec:3}
%The \"uber-gravity model suggest a cosmological model which is simple in the sense of the background dynamics.
In this section, we propose a cosmological model which is a natural consequence of \"uber-gravity model.
According to Fig. \ref{fig:trace-full}, we see that the \"uber-gravity leads to a very simple model for the gravity as
\begin{eqnarray} \label{uber-cosmology}
\text{Gravity} \simeq
\begin{cases}
\text{$R=R_0$} & \rho< \rho_{{\rm \ddot{u}ber}} \\
\text{$\Lambda$CDM}  & \rho > \rho_{{\rm \ddot{u}ber}}
\end{cases}
\end{eqnarray}
which we call \"u$\Lambda$CDM.  In this scenario, if matter density $\rho > \rho_{{\rm \ddot{u}ber}}$ then it sees pure GR with a cosmological constant, while if $\rho < \rho_{{\rm \ddot{u}ber}}$ then the metric is constrained to have constant Ricci scalar  i.e. $R_0$, which is a free parameter in our model presented in Eq.(\ref{ubergravity-Lagrangian}). We should mention that the above argument does not depend on the radiation content of the universe since the radiation is trace-free and has no contribution to our conclusion based on Fig. \ref{fig:trace-full}.% Furthermore, since our critical density is sensitive to the total energy density then even in radiation dominant era we satisfy $\rho>\rho_{{\rm \ddot{u}ber}}$ even if matter is not the dominant content.

The sharp transition in our model is representative of a family of models that have different physics for early and late time universe. Such models may address the tensions between early and late time observations. In this sense, our model (\ref{uber-cosmology}) is very similar to vacuum metamorphosis scenario \cite{Caldwell:2005xb} though they are conceptually different and we do not have any claim about the vacuum structure \cite{Sakharov:1967pk}.

In the following sections, we study the background and perturbation of this model.

\section{Background Analysis and CMB}
\label{Sec:b}
The continuity equation for matter gives $\rho(z)\propto (1+z)^{3}$ which means it is decreasing and the universe is in pure $\Lambda$CDM phase, i.e. $\rho>\rho_{{\rm \ddot{u}ber}}$ in (\ref{uber-cosmology}) for early times. Then there is a transition redshift $z_\oplus$ given by $\rho_{{\rm \ddot{u}ber}}$ when the model switches to $R=R_0$ phase in (\ref{uber-cosmology}).
For the background we assume a spatially flat FRW metric which gives the following (modified) Friedmann equation for $z>z_\oplus$
\begin{equation}\label{hubble-high}
E^2(z) = \Omega_{m}(1+z)^3 +\Omega_\Lambda,
\end{equation}
where $E(z)\equiv H(z)/H_0$ and $H_0$ is Hubble parameter at $z=0$. %Also note that $\Omega _{m,\oplus} = 8\pi G\rho_{{\rm \ddot{u}ber}} / 3H_0^2$.\\
For $z<z_\oplus$ we have
\begin{equation}
E^2(z) = \frac{1}{2}\bar{R}_0 +  (1-\frac{1}{2}\bar{R}_0)(1+z)^4
\end{equation}
where $\bar{R}_0\equiv R_0/6 H_0^2$. %In the following we will assume $\Omega_K^0\sim 0$ since it is compatible with our future parameter estimation and it is in agreement with the observations.
We assume $E(z)$ is continuous at $z=z_\oplus$ to read $\bar{R}_0$ from the following relation
\begin{equation}\label{cont-relation}
\Omega_{m}(1+z_{\oplus})^3+\Omega_\Lambda = \frac{1}{2}\bar{R}_0 + (1-\frac{1}{2}\bar{R}_0)(1+z_{\oplus})^4.
\end{equation}
Furthermore, we assume continuity of $H'(z)$ (prime is derivative wrt redshift) or equivalently Ricci scalar which gives us an additional constraint on our parameters
\begin{equation}\label{cont-relation1}
	\Omega_{m}(1+z_{\oplus})^3 = \frac{4}{3}\, (1-\frac{1}{2} \bar{R}_0 ) (1+z_{\oplus})^4.
\end{equation}
Therefore, we see that \"u$\Lambda$CDM has three independent free parameters i.e. $H_0$, $\Omega_m$, and $z_\oplus$  which is one more than standard $\Lambda$CDM's  $H_0$  and $\Omega_m$.
Now we are going to constrain \"u$\Lambda$CDM with observational data and contrast it with $\Lambda$CDM. To do a fair comparison, we should mention that in the following we will find the best fit of $\Lambda$CDM with exactly the same datasets which will be used for \"u$\Lambda$CDM. As such, the best fit values in $\Lambda$CDM may be slightly different from those of Planck 2015 \cite{Ade:2015xua}.

\subsection{Observational Datasets}
%We mainly focus on geometrical background data to constrain our free parameters e.g. same as \cite{Dhawan:2017leu}.
In the following we report the datasets used in this work including: CMB, local $H_0$, BAO and Lyman-$\alpha$ forest.

\begin{table*}[t]
	\begin{tabular}{ |c|c|c|c|c|c|c|c|c| }
		\hline
		CMB & BAO & BAO & Lyman-$\alpha$ & Hubble\\ \hline
		&  & & &  \\
		Planck 2015  \cite{Ade:2015xua} & 6dFGS ($z=0.106$) \cite{6df}  & LOWZ ($z=0.320$) \cite{Anderson:2013zyy}  & Ly$\alpha$ ($z=2.40$) \cite{Bourboux:2017cbm}& Local $H_0$ \cite{Riess:2016jrr}\\
		&  & & &  \\
		TT$+$lowP data	& $r_d/D_V=0.336\pm 0.015$ & $D_V = 1264.0\pm 25.0$ &$D_H/r_d=8.94 \pm 0.22$ &  $H_0 = 73.24 \pm 1.74 ~{\rm km/s/Mpc}$\\
		&  & & &   \\
		\hline
		&  & & &   \\
		 & MGS ($z=0.150$) \cite{Ross:2014qpa}  &CMASS ($z=0.570$) \cite{Gil-Marin:2015nqa} &Ly$\alpha$ ($z=2.40$) \cite{Bourboux:2017cbm} &  \\
		&  & & &  \\
			& $D_V = 664.0\pm 25.0$ & $D_V = 2056.0\pm 20.0$ & $D_M/r_d=36.6 \pm 1.2$& \\
		&  & & &  \\
		\hline
	\end{tabular}
	\caption{\label{tab:data}Datasets.}
\end{table*}

%For CMB, we focus on the position of the first peak which gives the distance between us and the surface of last scattering as one indicator \cite{Wang:2006ts,Elgaroy:2007bv,Komatsu:2008hk}. This distance is given by $100\Theta = 1.04085 \pm 0.00047$ and we need to use $r_* =   144.61 \pm 0.49$ Mpc and $z_* = 1090.09 \pm 0.42$ all reported by Planck 2015 TT$+$lowP data (see Table 4 in \cite{Ade:2015xua}).  We also add a data point as the value of  matter density parameter times the square of Hubble parameter from the CMB perturbations, $\Omega_m h^2=0.1415\pm 0.0019$ \cite{Ade:2015xua}. The reason for this is that \"u$\Lambda$CDM cannot affect the CMB anisotropy power spectrum (up to small secondary effects which are outside the scope of this analysis) since the behavior of \"u$\Lambda$CDM and $\Lambda$CDM are identical at $\rho > \rho_{{\rm \ddot{u}ber}}$ or high redshifts according to (\ref{hubble-high}). Therefore,  we encoded all CMB constraints into two main data points, i.e. distance to the last scattering surface and matter density.% In the following, by CMB we meant these two data points.
For CMB, we use the Planck 2015 TT$+$lowP data \cite{Ade:2015xua}.
Another data point is given by Riess et al. \cite{Riess:2016jrr} i.e. $H_0 = 73.24 \pm 1.74$ km/s/Mpc, and from now on we refer to it by R16. This is the data point which is in tension with Planck 2015 best-fit $\Lambda$CDM model. %Consequently, we will not use other SNe datasets which may need to consider the correlation between different samples and it is beyond scope of this work.
The other dataset is the baryonic acoustic oscillation (BAO) measurements: we use the  6dFGS data at $z = 0.106$ \cite{6df}, the SDSS main galaxy (MGS) at $z = 0.15$ of \cite{Ross:2014qpa}, Baryon Oscillation Spectroscopic Survey (BOSS) LOWZ \cite{Anderson:2013zyy} at $z= 0.32$, and CMASS surveys \cite{Gil-Marin:2015nqa} at $z=0.57$.
%In addition to the above datasets
To test our model, we consider BAO in Lyman-$\alpha$ forest of quasar spectra by \cite{Bourboux:2017cbm} who report two independent quantities at $z=2.40$; line of sight distance as $D_H/r_d=8.94 \pm 0.22$ and angular distance as $D_M/r_d=36.6 \pm 1.2$ where $D_M=(1+z)\,D_A$.
A tension between Planck 2015 and Lyman-$\alpha$ forest BAO has been reported \cite{Font-Ribera:2013wce} which could potentially be solved with a dynamical dark energy \cite{Zhao:2017cud}. Here in this work we use the recent analysis \cite{Bourboux:2017cbm} which has less tension with Planck 2015. We summarized the datasets in Table \ref{tab:data}.
In the following, we will report the best fit of our model and standard $\Lambda$CDM, with CMB+BAO+R16 which makes our results comparable with Planck 2015 \cite{Ade:2015xua}. %Then we will add two Lyman-$\alpha$ BAO constraints to check if \"u$\Lambda$CDM can lessen the reported tension.

\subsection{Results}

\begin{figure*}
	\centering
	\includegraphics[width=0.8\linewidth]{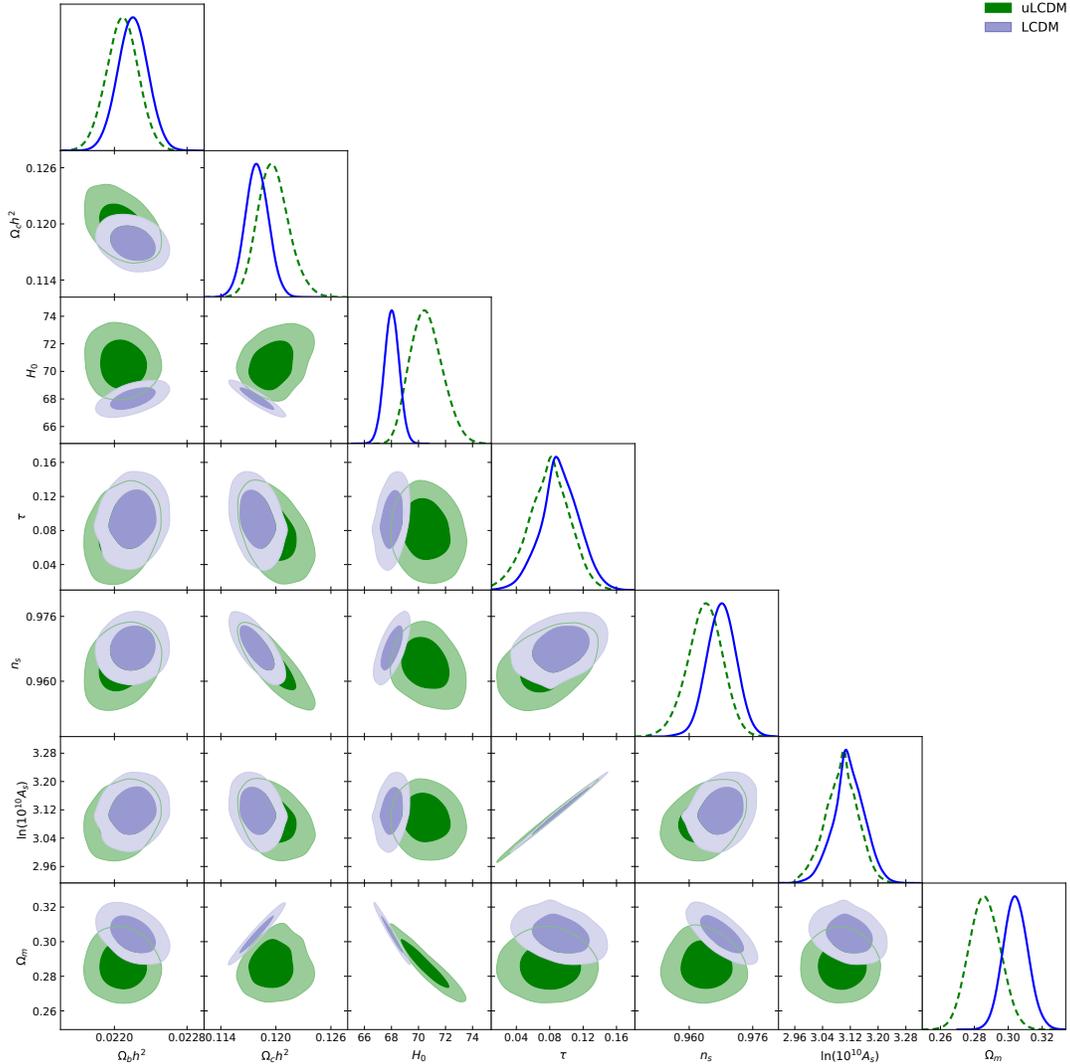}
	\caption{Comparison of the base $\Lambda$CDM model with   \"u$\Lambda$CDM  parameter constraints from data set Table \ref{tab:data}.
 It is obvious that \"u$\Lambda$CDM prefers higher $H_0$ and less $\Omega_m$ in comparison with $\Lambda$CDM.}
	\label{fig:uLCDM-LCDM}
\end{figure*}

\begin{figure*}
	\centering
	\includegraphics[width=0.8\linewidth]{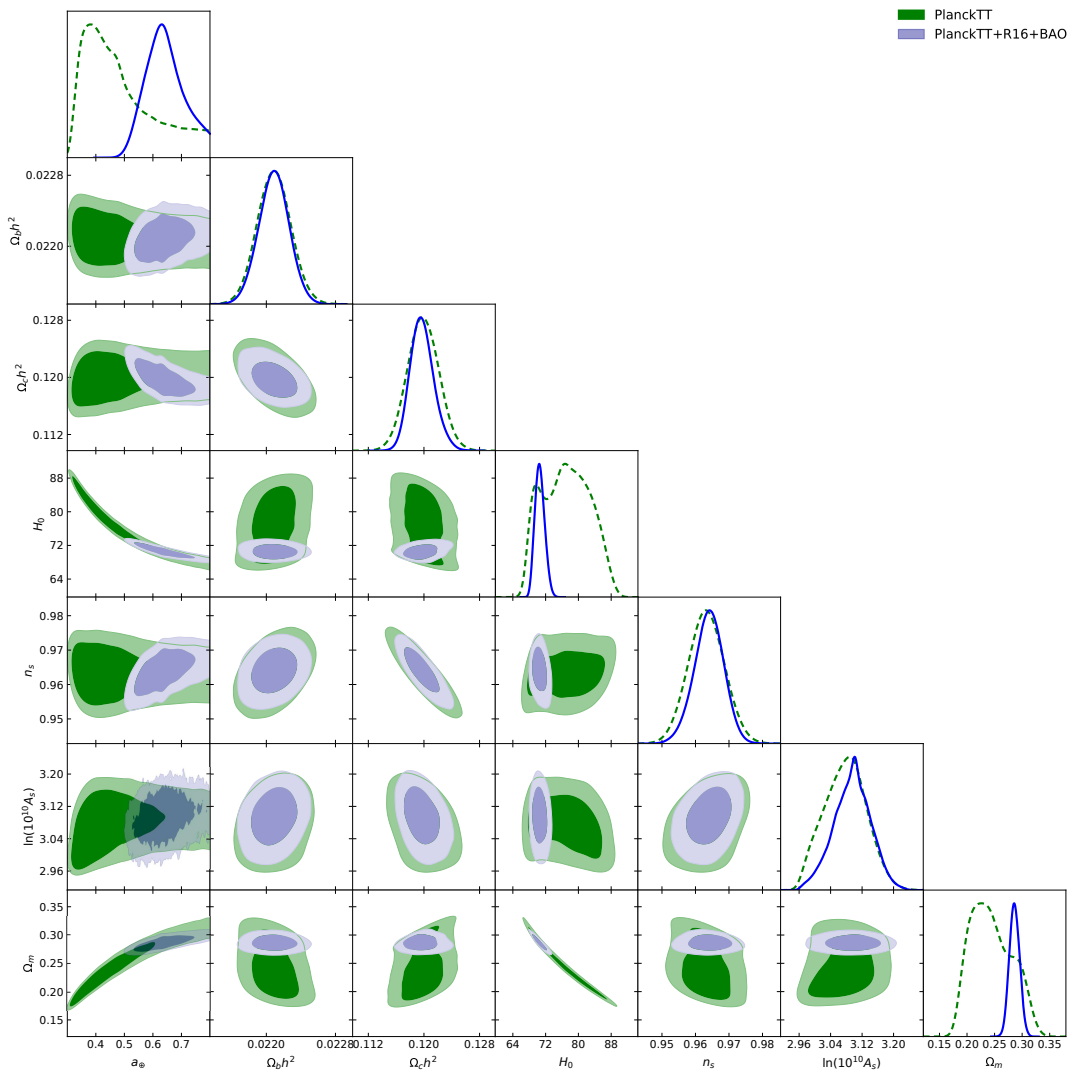}
	\caption{The comparison of  \"u$\Lambda$CDM  with base CMB temperature and polarization data (green contour plots) and CMB+BAO+R16 constrained (gray contour plots)}
	\label{fig:uLCDM}
\end{figure*}

Our results are summarized in Table \ref{tab:best-fit}. %We used two different combinations of datasets: First we use the data used in Planck 2015 \cite{Ade:2015xua},  and the second combination is Planck data with the addition of  BAO data points and SNeIa R16 data set.
We contrast best-fit parameters and goodness of fit between \"u$\Lambda$CDM and standard $\Lambda$CDM with these datasets \footnote{Note that the best-fit values of $\Lambda$CDM may differ slightly from Planck 2015 \cite{Ade:2015xua} due to simplified analysis and different dataset combinations.}. Also we conduct Akaike information criterion (AIC) to compare the two model with the data set.
In Figure (\ref{fig:uLCDM-LCDM}), we plot the confidence level of standard $\Lambda$CDM model in comparison with \"u$\Lambda$CDM, the data set used is CMB+R16+BAO.  In Figure (\ref{fig:uLCDM}), we plot the contour plot of the free parameters of \"u$\Lambda$CDM and illustrated the best fits graphically for CMB data set and CMB+BAO+R16.  For $\Lambda$CDM  the best fit values for derived parameters of $\Omega_m=0.3044 \pm {0.0073}$ and $H_0=68.02 \pm 0.55$ km/s/Mpc with $\chi^2=923.3$ . A little bit higher value for $H_0$ in comparison with Planck 2015 \cite{Ade:2015xua} is because we have added R16 to our data set which drives a higher value for Hubble parameter (the rest of parameters are listed in Table \ref{tab:best-fit}). Our model best fit occurs at $H_0=70.6_{-1.3}^{+1.1}$ km/s/Mpc, $\Omega_m=0.2861 \pm {0.0092} $ and the transition scale factor  $a_{\oplus}= (1+z_{\oplus})^{-1} =0.642_{-0.078}^{+0.056}$ with $\chi^2=921.9$. Since \"u$\Lambda$CDM and $\Lambda$CDM don't have the same number of free parameters then $\chi^2$ analysis may not be very useful. Because of it we have used Akaike information criterion (AIC) analysis \cite{AIC} which is basically a simplified version of Bayesian analysis and it takes care of the number of free parameters. The AIC results show \"u$\Lambda$CDM is slightly preferred by the datasets even with one more free parameter.

\begin{figure}
	\centering
	\includegraphics[width=1\linewidth]{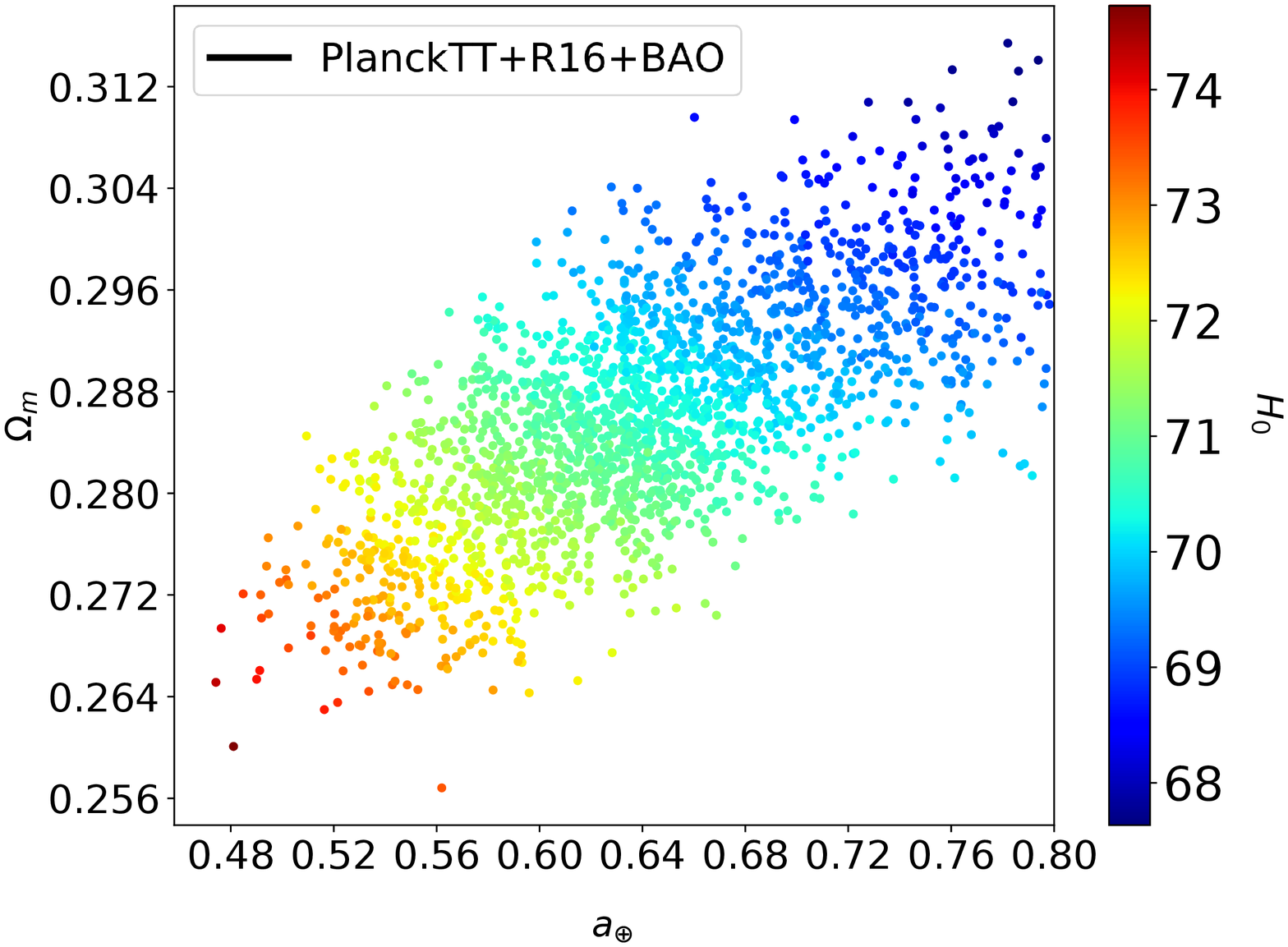}
	\caption{A 3D plot of $\Omega_m$ versus the transition scale factor $a$ and Hubble parameter.}
	\label{fig:3Dplot}
\end{figure}

\begin{figure}
	\centering
	\includegraphics[width=1\linewidth]{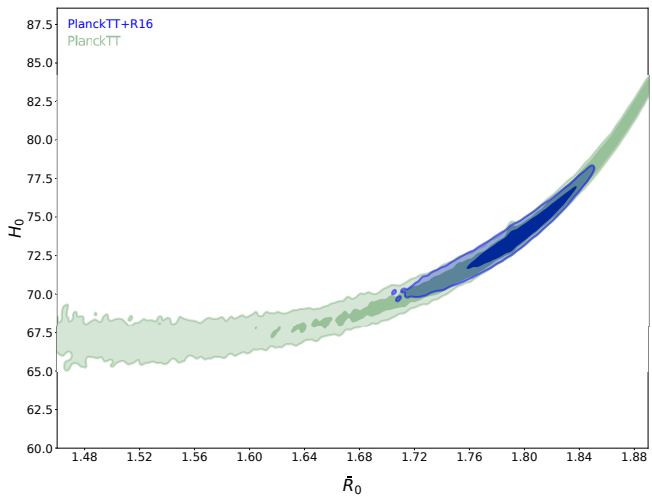}
	\caption{The confidence level of $R_0$ and $H_0$ parameter.}
	\label{fig:H0-R0}
\end{figure}

In Fig. \ref{fig:3Dplot}, we show the confidence level of $\Omega_m$, $a_{\oplus}$ and the color-bar showed the value of $H_0$. This plot also indicates the anti-correlation of the matter density and Hubble parameters, the data set which is used in this plot is CMB+R16+BAO. In Fig. \ref{fig:H0-R0}, we have the contour plot of $H_0$ and free parameter $\bar{R}_0$. This is a crucial plot to compare our results with \cite{DiValentino:2017rcr} (see Figure(4) in their paper) which shows the consistency of two models. Green contours shows the constrains from CMB data alone and blue contours obtained from CMB+R16.

\begin{figure}
	\centering
	\includegraphics[width=1\linewidth]{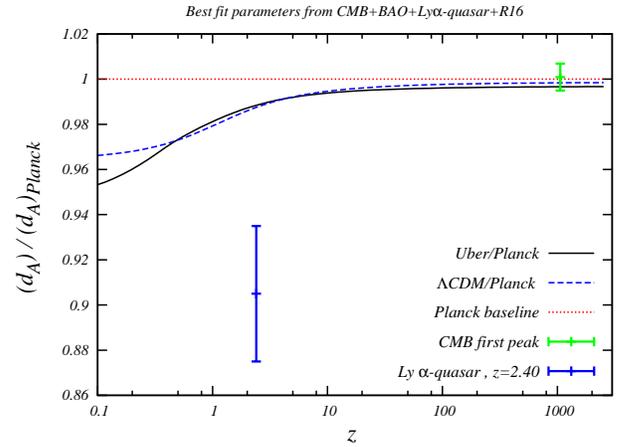}
	\caption{The angular diameter distance $d_A(z)$ (normalized to Planck 2015 best fit values' prediction) for \"u$\Lambda$CDM model and the best fit for $\Lambda$CDM has been plotted in solid black and dashed blue lines, respectively. The data point at $z=1090$ is the distance of last scattering surface given by $d_A(z) = \frac{r_s}{(1+z)\Theta}$ and $100\, \Theta= 1.04085\pm\,0.00047$ reported by Planck 2015 \cite{Ade:2015xua}. %Obviously our model can satisfy this very tight constraint perfectly.
	We also added the BAO angular distance  from Lyman-$\alpha$ quasar  $D_M/r_d=36.6 \pm 1.2$ reported in \cite{Bourboux:2017cbm}.}
	\label{fig:chiMrel}
\end{figure}

\begin{figure}
	\centering
	\includegraphics[width=1\linewidth]{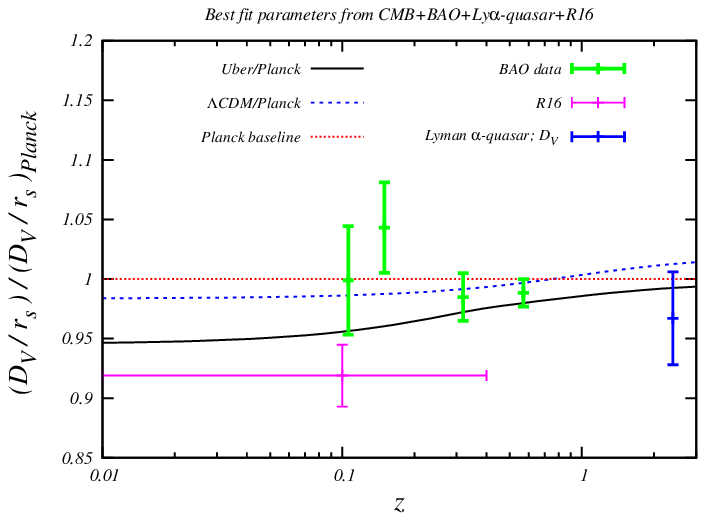}
	\caption{The volume distance, $D_V(z)$ (normalized to Planck 2015 best fit values' prediction) for our model and best fit of $\Lambda$CDM  are plotted in solid black and dashed blue lines, respectively. Green data points represents four BAO measurements. In addition we added BAO data point from Lyman-$\alpha$ quasar in dark blue. We (schematically) translated R16 measurement for $H_0$ to a volume distance which is the blue data point. Obviously, \"u$\Lambda$CDM can decrease $H_0$ tension significantly while it is still compatible with BAO data points.}
	\label{fig:chiVrel}
\end{figure}

We summarize the results in Table \ref{tab:best-fit} and based on these values we plot background distance quantities: In Fig. \ref{fig:chiMrel}, the angular diameter distance normalized to Planck 2015 best fit values has been plotted. In addition to our distance to the last scattering surface, we plotted the Lyman-$\alpha$ forest BAO data point at $z=2.4$ which shows a 2.5$\sigma$ tension with both models. %As mentioned above Lyman-$\alpha$ BAO is in tension with Planck 2015. $\Lambda$CDM seems to lessen this tension but it cannot resolve $H_0$ tension as we will see in Fig. \ref{fig:chiVrel}. However our model reduces this tension while it is compatible with CMB first peak data.
In Fig. \ref{fig:chiVrel}, we plotted the volume distance normalized to Planck 2015 best fit values. We have added BAO data and one should compare this plot with Fig. 14 in \cite{Ade:2015xua}. In addition to BAO we transformed local $H_0$ measurement \cite{Riess:2016jrr} to a distance. Planck 2015 and our best-fit $\Lambda$CDM model are in tension with R16. However the tension almost disappears in \"u$\Lambda$CDM model, while the tension with Lyman-$\alpha$ $D_V(z)$ measurement is reduced.

% *****************************************************

% *************************NEW PLOT ****************************

% *****************************************************

% *************************NEW PLOT ****************************

% *****************************************************

% *************************NEW PLOT ****************************

% *****************************************************

%In Figure (\ref{fig:H0-R0}), we plot the confidence level of the $H_0$ parameter with $\bar{R}_0$ in order to compare our results with the findings of DiValentino et al. \cite{DiValentino:2017rcr}, which is in a very good agreement.

%
%
%\begin{table}
%	\begin{tabular}{ |c|c|c| }
%		\hline
%		& $\Lambda$CDM & \"u$\Lambda$CDM \\ \hline
%		\multirow{4}{*}{\rotatebox[origin=c]{90}{ CMB+BAO+R16 }} & $\chi^2=7.84$ & $\chi^2=4.93$ \\
%		&  &  \\
%		& $\Omega_m=0.293^{+0.007}_{-0.009}$ & $\Omega_m=0.278^{+0.014}_{-0.011}$ \\
%		&  &  \\
%		& $H_0=69.4^{+0.8}_{-0.6}$ & $H_0=71.5_{-1.7}^{+1.6}$ \\
%		&  &  \\
%		&  & $z_\oplus=0.537_{-0.375}^{+0.277}$ \\ \hline
%		\multirow{3}{*}{\rotatebox[origin=c]{90}{ + Lyman-$\alpha$ BAO }} & $\chi^2=19.25$ & $\chi^2=17.11$ \\
%		&  &  \\
%		& $\Omega_m=0.289^{+0.008}_{-0.007}$ & $\Omega_m=0.280\pm 0.013$ \\
%		&  &  \\
%		& $H_0=69.8^{+0.7}_{-0.8}$ & $H_0=71.2^{+1.8}_{-1.5}$ \\
%		&  &  \\
%		&  & $z_\oplus=0.413^{+0.305}_{-0.289}$ \\
%		\hline
%	\end{tabular}
%	\caption{\label{tab:best-fit}The best fit values for $\Lambda$CDM and \"u$\Lambda$CDM for two sets of data.}
%\end{table}

\begin{table}
	\begin{tabular}{ |c|c|c| }
		\hline
		& $\Lambda$CDM & \"u$\Lambda$CDM \\ \hline
		\multirow{14}{*}{\rotatebox[origin=c]{90}{  Main Parameters  }}
        &  &  \\
        & $\Omega_c h^2 = 0.1179 \pm 0.0013 $ & $\Omega_c h^2 = 0.1197 ^{+0.0015}_{-0.0018}  $ \\
		&  &  \\
        & $\Omega_b h^2 = 0.0222 \pm 0.002 $ & $\Omega_b h^2 = 0.0221 \pm 0.0002  $ \\
		&  &  \\
        & $\Omega_\Lambda = 0.690 ^{+0.0070} _{-0.0075}  $ & $\Omega_\Lambda = 0.7139 \pm 0.0092  $ \\
		&  &  \\
        & $\tau = 0.092\pm 0.023$ & $\tau = 0.079 \pm 0.024 $ \\
		&  &  \\
        & $\ln(10^{10} A_s) = 3.116^{+0.047}_{-0.041}$ & $\ln(10^{10} A_s) = 3.094 \pm 0.047 $ \\
		&  &  \\
        & $n_s = 0.9681 \pm 0.0037 $ & $n_s = 0.9640 ^{+0.0046}_{-0.0041} $ \\
		&  &  \\
        & ----- & $z_\oplus=0.537_{-0.375}^{+0.277}$ \\
       &  &  \\
        \hline
        &  &  \\
        & $\Omega_m= 0.3044 \pm 0.0073$ & $\Omega_m = 0.2861 \pm 0.0092$ \\
		&  &  \\
		& $H_0=68.02\pm 0.55$ & $H_0= 70.6 ^{+1.1} _{-1.3}$ \\
		&  &  \\
		\hline
		\multirow{5}{*}{\rotatebox[origin=c]{90}{ Stat. }}
		&  &  \\
		& $\chi^2=923.3$ & $\chi^2= 921.9$ \\
		&  &  \\
				& $AIC = 1858.6$ & $AIC = 1857.8$ \\
		&  &  \\		
		\hline
	\end{tabular}
	\caption{\label{tab:best-fit}The best fit values for $\Lambda$CDM and \"u$\Lambda$CDM for two sets of data. The Akaike information criterion (AIC) analysis shows \"u$\Lambda$CDM is slightly better than $\Lambda$CDM.}
\end{table}

% *****************************************************
% *****************************************************

% *****************************************************
% *****************************************************

%%%%%%%%%%%%%%%%%%%%%%%%% Perturbations %%%%%%%%%%%%%%%%%%%%%%%%%%%%%%%%%%%%%%
%%%%%%%%%%%%%%%%%%%%%%%%%%%%%%%%%%%%%%%%%%%%%%%%%%%%%%%%%%%%%%%

\section{Perturbations: the set-up}\label{pert}
It is well-known that all $f(R)$ theories of gravity can be written as scalar-tensor theories. Consider the following scalar-tensor action representing the cosmological era after the transition in \"uber-gravity:
\begin{equation}
S=\frac{1}{16\pi G}\int d^{4}x\sqrt{-g}\,\bigg[ \xi \, (R-R_0)-\lambda\bigg]+{\cal{L}}_{m}\,, \label{action}
\end{equation}
where $R$ is the Ricci scalar, $g$ is the trace of the metric $g_{ab}$, $R_0$ is the (constant) value of curvature after the transition, and $\xi$ is a Lagrange multiplier that is a priori space-time dependent and ensures the constraint $R=R_0$. In addition we know from the physics that the above action should be matched with standard $\Lambda$CDM which means for $\xi=1$ it should be standard Einstein-Hilbert action i.e. we should set $\lambda=2\Lambda-R_0$ where $\Lambda$ is the cosmological constant for $ \rho > \rho_{{\rm \ddot{u}ber}}$. The equations of motion (EOM) for this action are:
\begin{eqnarray}\nonumber
-\frac{ g_{ab}}{2}\xi (R\!-\!R_0)\!+\!\frac{\lambda}{2} g_{ab}\!+\!\xi R_{ab}\! -\![\nabla_{a}\!\nabla_{b}\!-\!g_{ab}\Box] \xi \!\!\!&=& \!\!\!8 \pi G T_{ab} \\\label{eom1}\\
R- R_0\!\!\!&=&\!\!\!0\,. \label{eom2}
\end{eqnarray}
The trace of Eq.~\eqref{eom1} can be written using the constraint equation as:
\begin{equation}
\xi R_0=8 \pi G T-2\,\lambda-3\,\Box \xi\,, \label{eom3}
\end{equation}
where $T=g^{ab}T_{ab}$. Using the results for the linear scalar perturbation of Appendix ~\ref{secperturbations} we can write the Newtonian potential $\psi$  and lensing potential $\phi_L$ in the quasi-static regime ($\nabla^2 \gg {\cal H}^2$):
\begin{eqnarray}
\nabla^2\psi=\frac{16\pi G a^2}{3 \xi^0}\delta\rho,\\ \label{poisson}
\phi_L\equiv \frac{\phi+\psi}{2} = \frac{3}{4} \psi, \\
\left[\dot{\psi}+\left({\cal H}+\frac{\dot{\xi^0}}{\xi^0}\right)\psi\right]_{,i} = -\frac{16 \pi G}{3 \xi^0} \bar{\rho} u_i,  \label{momentum}
\end{eqnarray}
where $\bar{\rho}(\tau)+\delta\rho({\bf x},\tau)$ and $u_i({\bf x},\tau)$ are the CDM density and peculiar velocity, respectively.

\begin{figure}
	\centering
	\includegraphics[width=1\linewidth]{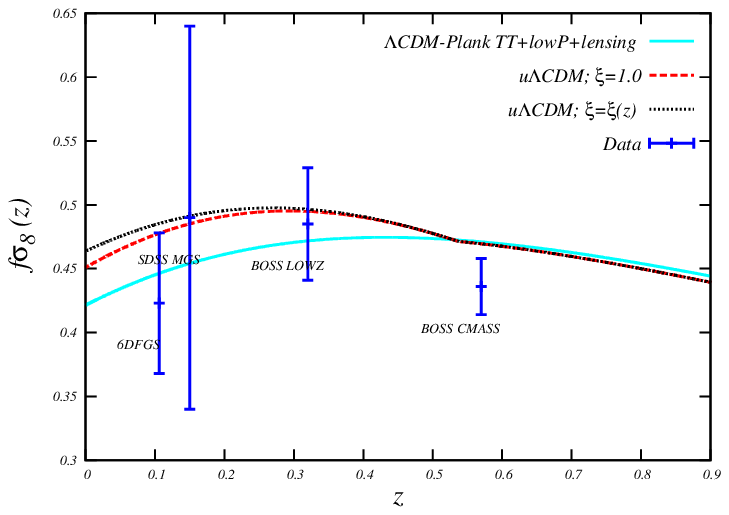}
	\caption{The growth rate is plotted versus redshift for $\Lambda$CDM model with the best fit from Planck data (TT+lowPlensing best fit) and \"u$\Lambda$CDM with best fit from (CMB+BAO+R16). The data points are growth rate times $\sigma_8$ from 6DFGS + SDSS-MGS + BOSS-Lowz + BOSS-CMASS. In \"u$\Lambda$CDM case we assumed both constant $\xi=1$ and evolving $\xi(z)$ for two sets of best fit parameters in TABLE \ref{tab:best-fit}.}
	\label{fig:growthrate}
\end{figure}

\begin{figure}
	\centering
	\includegraphics[width=1\linewidth]{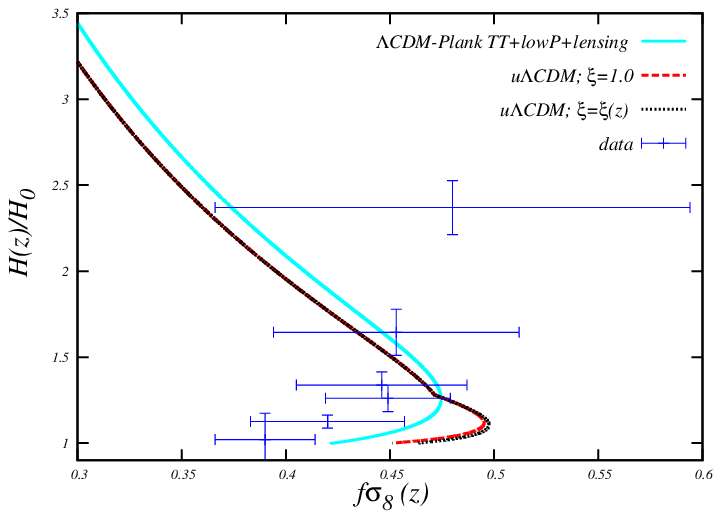}
	\caption{The normalized growth factor is plotted versus $f\sigma_8$. The  solid line indicate the prediction of standard $\Lambda$CDM while \"u$\Lambda$CDM has been plotted for different scenarios similar to FIG. \ref{fig:growthrate}. The data points are from \cite{Moresco:2017hwt}.}
	\label{fig:fE}
\end{figure}

In order to solve Equations (\ref{eom3}-\ref{momentum}), we first need to know the initial conditions for the fields at $z=z_\oplus$. Comparing the action (\ref{action}) with the Einstein-Hilbert action, we find:
\begin{equation}
\xi^0(z \geq z_\oplus)=1~~{\rm and} ~~ \dot{\xi}^0(z \geq z_\oplus)= 0,
\end{equation}
which sets the initial condition for the background equation for $\xi^0$ in (\ref{eom3}). Having solved for $\xi^0(\tau)$, we can plug into Equation (\ref{poisson}) to find Newtonian potential, which in turn governs the geodesic equation for CDM. By comparing Equations (\ref{poisson}-\ref{momentum}) with Einstein equation, we notice that continuity of matter density and velocity implies that there will be a jump in Newtonian potential, while the lensing potential will remain continuous at $z=z_\oplus$:
\begin{eqnarray}
\psi(z < z_\oplus) &=& \frac{4}{3} \psi(z > z_\oplus), \\
\phi_L(z < z_\oplus) &=& \phi_L(z > z_\oplus), \\
{\rm as} ~ z & \rightarrow & z_\oplus.
\end{eqnarray}
Therefore, the rate of structure formation (at the linear level), which is governed by gravitational acceleration, suddenly jumps by 33\% at the onset of the transition.
In order to quantify the growth of the structures in linear regime we have to determine the evolution of the growth rate parameter $f\equiv d\ln \delta / d\ln a$ (which is the logarithmic derivative of dark matter density contrast with respect to scale factor). This evolution is governed by the continuity and Euler equation and also the modified Poisson equation which results in
\begin{equation}
\frac{df}{dz}+[\frac{d\ln E(z)}{dz} - \frac{2}{1+z}]f - \frac{f^2}{1+z} + \frac{2\Omega_m (1+z)^2}{E^2(z)\xi(z)} =0,
\end{equation}
where we should note that the $\xi$ has a dynamic determined from field Einstein field equation as
\begin{equation}\label{xi}
\frac{d\xi}{dz} + [\frac{1}{1+z} - \frac{d\ln E(z)}{dz}]\xi = - \frac{\Omega_m (1+z)^2}{E^2(z)} + \frac{\bar{R}_0 - \Omega_{\Lambda}/2}{(1+z)E^2(z)}
\end{equation}
In Fig.(\ref{fig:growthrate}) we plot the growth rate versus redshift for $\Lambda$CDM and  \"u$\Lambda$CDM for two sets of parameters  reported in TABLE \ref{tab:best-fit}. In addition, we assumed that both constant $\xi=1$ and evolving $\xi(z)$ (which satisfies (\ref{xi})) while the background cosmology is governed by \"u$\Lambda$CDM.
% *****************************************************
% *****************************************************
In Fig.(\ref{fig:fE}), we plot the growth rate versus dimensionless Hubble parameter, this plot probe a cosmology with expansion history of the Universe and growth rate of perturbations proposed by Linder \cite{Linder:2016xer}. The data points of expansion history versus growth rate is from Moresco and Marulli \cite{Moresco:2017hwt}.

% *****************************************************
% *****************************************************

The real story, of course, is more complicated. Nonlinear structures are already well in place by $z_\oplus \sim 0.4$. Inside haloes and their outskirts, the density never goes below $\rho_{{\rm \ddot{u}ber}}$, implying that GR remains valid. The voids, however, could have underdensities of $\sim 50\%$, and thus have crossed over in the \"uber-era, much earlier \footnote{This is a particular extreme of the well-known Chameleon screening \cite{Khoury:2003aq}, which is ubiquitous for generic scalar-tensor theories.}. The boost in Newtonian potential can accelerate the emptying of the voids and boost the Integrated Sachs-Wolfe (ISW) effect. Could this provide a means to understand the void phenomenon \cite{Peebles:2001nv}, or the anomalously large ISW effects observed  in voids \cite{Kovacs:2017hxj} and in general \cite{Ho:2008bz}? We defer studying these possibilities to future work, but comment that, due to their nonlinear nature, they can only be satisfactorily addressed using numerical simulations.

%In order to study scalar perturbations in this model we consider the following metric:
%\begin{equation}
%ds^2=a^2[(1+2 \,\epsilon \,\psi)d\tau^2+(1-2\,\epsilon\, \phi)d{\text{\bf{x}}}^3]\,,\label{pertmetric}
%\end{equation}
%where $a$ is the scale factor and $\epsilon$ is the perturbation parameter. With this notation a perturbation in the lagrange multiplier reads $\xi=\xi^0+\epsilon \,\xi^1$ and the Ricci scalar $R=R^0+\epsilon \,R^1$. Note Eq.\eqref{eom2} implies that $R^1=0$ and then $R=R^0=R_0$ in addition to:
%\begin{equation}
%R^1=\frac{2}{a^2}(2\nabla^2\phi-\nabla^2 \psi)=0\,. \label{riccipert}
%\end{equation}
%On the other hand, the perturbation of the trace of Eq.\eqref{eom1}, the temporal component of the Einstein tensor and the sum of the spatial component of the Einstein tensor give:
%\begin{eqnarray}
%\xi ^1R_0\!\!&=&\!\!8 \pi G T^1-3\frac{\nabla^2\xi^1}{a^2}\,, \label{tracepert} \\
%\nabla^2 \phi \!\!&=&\!\!-\frac{4 \pi G a^2 \delta T_0^0} {\xi^0}+\frac{\nabla^2 \xi^1}{2\xi^0}+\frac{R_0 a^2\xi^1}{4 \xi^0}\,, \label{00pert}\\
%\nabla^2(\psi-\phi)\!\!&=&\!\!\frac{4 \pi G}{\xi^0}a^2\sum_{i}\delta T_i^i-\frac{\nabla^2\xi^1}{\xi^0}-\frac{3\xi^1}{4\xi^0}R_0a^2, \label{iipert}
%\end{eqnarray}
%where we have $T_{ij}=\bar{T}_{ij}+\epsilon \, \delta T_{ij}$. With Eqs. \eqref{00pert},\eqref{iipert} we can find the equation for the lensing potential
%\begin{equation}
%\nabla^2()
%\end{equation}
%
%The trace of Eq.~\eqref{eom1} is

%%%%%%%%%%%%%%%%%%%%%%%%%%%%%%%%%%%%%%%%%%%%%%%%%%%%%%%%%%%%%%%
%%%%%%%%%%%%%%%%%%%%%%%%%%%%%%%%%%%%%%%%%%%%%%%%%%%%%%%%%%%%%%%

\section{Concluding remarks}
\label{Sec:4}
In this work, we show how from the idea of \"uber-gravity a cosmological model is emerged. We call this model
\"u$\Lambda$CDM, to indicate two distinct phases of cosmological evolution: The era of $\Lambda$CDM,
and the \"uber-era with a constant Ricci scalar. The universe is in pure $\Lambda$CDM and GR when matter density is larger than a critical density, $\rho_{\rm \ddot{u}ber}$. After matter density drops below $\rho_{\rm \ddot{u}ber}$, the universe is in a state with a constant Ricci scalar where we find a suitable solution for Hubble parameter to match the data. This behavior can be seen in a more general context, as a phase transition in gravity, and \"uber-gravity, naturally, provides such a framework to think about such a phase transition.
We showed, at the level of background, \"u$\Lambda$CDM can be a potential resolution for the tension between high and low redshift $H_0$ measurements, noting that the $H_0$ measured in local universe is computed in the \"uber-era. We also show that in the level of background the \"u$\Lambda$CDM model fits with the BAO data better than $\Lambda$CDM, albeit marginally.

Furthermore, we provide a preliminary analysis of structure formation in \"u$\Lambda$CDM, showing that structure formation will be enhanced in the \"uber-era. This is most likely to affect cosmic voids, and could potentially explain anomalies associated with void structure formation. We plan to study this possibility in the future.

{\it{Note-I}}: Recently, LIGO reported detection of gravity wave from a NS-NS binary with its EM counterpart \cite{ligo}. By using gravitational wave as standard siren (which is completely independent of SNe or CMB) they could measure Hubble parameter, $H_0=70.0^{+12.0}_{-8.0}$ km/s/Mpc \cite{Abbott:2017xzu}, which as of yet cannot distinguish the models discussed here. Higher statistics of such observations can reduce the errors and shed light on the status of $H_0$ tension in cosmology.

{\it{Note-II}}: During the final stages of this work, Valentino, Linder and Melchiorri submitted a preprint that addressed the $H_0$ tension via Parker's model of Vacuum Metamorphosis (VM) \cite{DiValentino:2017rcr} which has a very similar structure to our model. The difference is that they also consider other cosmological parameters beyond $\Lambda$CDM to improve the fit, while they do not provide a consistent treatment for perturbations in VM.

\vspace{.5 cm}
\textit{Acknowledgments:}
We are grateful to P. Creminelli, A.-C. Davis, Mohammad Ali Islami, J. Khoury,  Michele Ennio Maria Moresco, S. Rahvar, M. M. Sheikh-Jabbari and  R. K. Sheth for insightful comments and discussions. We should thank Hossein Mos'hafi for his extensive discussions on the CMB data analysis part. We also thank the anonymous referee for her/his valuable comments.
SB and NK would like to thank NORDITA workshop on ``Advances in theoretical cosmology in light of data", where this work was initiated there and the Abdus Salam International Center of Theoretical Physics (ICTP) for a very kind hospitality, which the main part of this work has been done there. NK also thanks Perimeter Institute, CITA and Orsay/Saclay (``DarkMod workshop/conference") for their supports during completion of this work. This research is partially supported by Sharif University of Technology Office of Vice President for Research under Grant No. G960202. NK would like to 
thank the research council of Shahid Beheshti University 
for their supports.  \\ \\

 \appendix
 \section{Perturbations} \label{secperturbations}
In this section, we derive the background and linearly perturbed EOMs of Eqs.~\eqref{eom1}-\eqref{eom3}. To this end, we expand the EOM to linear order in scalar metric perturbations in the longitudinal gauge:
\begin{eqnarray}
ds^2\!\!&=&\!\!a^2(\tau)[-(1+2 \psi)d\tau^2+(1-2  \phi)d^3{\text{\bf{x}}}]\,,\label{pertmetric} \\
R \!\!&=&\!\! R_0 \,, \label{pertR} \\
\xi \!\!&=&\!\!\xi^0(\tau)+ \xi^1({\bf x},\tau) \label{pertxi} \\
T_{ab}\!\!&=&\!\!\bar{T}_{ab}(\tau)+\delta T_{ab}({\bf x},\tau)\,, \label{pertTab}\\
T\!\!&=&\!\!\bar{T}(\tau)+\delta T({\bf x},\tau)\,, \label{pertT}
\end{eqnarray}
where $T=T_{ab}g^{ab}$. The background EOMs are:
\begin{eqnarray}
\frac{R_0 a^2}{3}\!\!&=&\!\! 2\frac{\ddot a}{a} \,,\label{0pertR}\\
%\frac{R_0 a^2}{3}\!\!&=&\!\! \frac{8\pi G \bar{T} a^2}{3\xi^0}-\frac{\Box \xi^0}{\xi^0}\tcr{+}2 {\cal{H}}\frac{\dot{\xi^0}}{\xi^0}\,, \label{0perttrace} \\
\frac{R_0 a^2}{3}\!\!&=&\!\! \frac{8\pi G \bar{T} a^2}{3\xi^0}+2 {\cal{H}}\frac{\dot{\xi^0}}{\xi^0} + \frac{\ddot{\xi^0}}{\xi^0} -\frac{2}{3}a^2\lambda \,, \label{0perttrace} \\
%\frac{R_0 a^2}{3}\!\!&=&\!\! \frac{16\pi G \bar{T}^0_0 a^2}{3\xi^0}+\tcr{2}{\cal{H}}({\cal{H}}+\frac{\dot{\xi}^0}{\xi^0})-\frac{2}{3}\frac{\nabla^2 \xi^0}{\xi^0}\,, \label{0pertG00}
\frac{R_0 a^2}{3}\!\!&=&\!\! \frac{16\pi G \bar{T}^0_0 a^2}{3\xi^0}+2{\cal{H}}({\cal{H}}+\frac{\dot{\xi}^0}{\xi^0}) -\frac{\lambda a^2}{3\xi^{(0)}}\,, \label{0pertG00}
\end{eqnarray}
where ${\cal{H}}=\dot{a}/ a$. \\
%Combining Eq.~\eqref{0perttrace} with \eqref{0pertG00} we have:
%\begin{equation}
%16 \pi G a^2 (\bar{T}_0^0-2\bar{T})+4 {\cal{H}}\dot{\xi}^0+6 {\cal{H}}^2\xi^0-3\ddot{\xi}^0=0\,. \label{comb0pert}
%\end{equation}
%With the background equations we can solve for $a, R_0$ and $\xi^0$.\\
%For the liner perturbations note that Eq.\eqref{eom2} implies that $R^1=0$ and then $R=R^0=R_0$.

The EOMs at linear order are:

\begin{widetext}
\begin{eqnarray}
\nabla^2(\psi-2\phi)\!\!&=&\!\! -\psi R_0 a^2-3{\cal{H}}(3\dot{\phi}+\dot{\psi})-3\ddot{\phi}\,,\label{1pertR} \\
\frac{R_0a^2}{3}\xi^1\!\!&=&\!\!\frac{8\pi G a^2}{3}\delta T-2{\cal{H}}(2\psi\dot{\xi}^0-\dot{\xi}^1)-\Box \xi^1-\dot{\xi}^0(3\dot{\phi}+\dot{\psi})-2\phi \ddot{\xi}^0\,, \label{1perttrace}\\
\frac{R_0a^2}{3}\xi^1\!\!\!&=&\!\!\!\frac{16\pi Ga^2}{3}\delta T_0^0\!+\!3\xi^1{\cal{H}}^2\!+\!\frac{4}{3}\xi^0\nabla^2\phi\!-\!{\cal{H}}({\cal{H}}\psi\!+\!\dot{\phi})\!-\!2{\cal{H}}(2\phi\dot{\xi}^0-\dot{\xi}^1)\!-\! \frac{2}{3}\nabla^2\xi^1\!.\label{1pertG00}
\end{eqnarray}
%We can read from the above equations the EOM for the Newtonian potential:
%\begin{equation}
%\nabla^2\phi=-\frac{4\pi G a^2}{\xi^0}(\delta T_0^0-2\delta T)+3{\cal{H}}({\cal{H}}\psi+\dot{\phi})+\frac{\dot{\xi}^0}{\xi^0}(2\psi {\cal{H}}-\frac{1}{4}\dot{\psi})-\frac{3}{2}\frac{\ddot{\xi}^0}{\xi^0}-\frac{3}{2}\frac{\xi^1}{\xi^0}{\cal{H}}^2-{\cal{H}}\frac{\dot{\xi}^1}{\xi^0}-\frac{3}{4}\frac{\Box \xi^1}{\xi^0}+\frac{1}{2}\frac{\nabla \xi^1}{\xi^0}
%\end{equation}
The $\{0i\}$ component of the equation of motion is
\begin{equation}
2\xi^0({\cal{H}}\psi_{,i}+\dot{\phi}_{,i})=8\pi G \delta T_{0i}-{\cal{H}}\xi^1_{,i}-\psi_{,i}\dot{\xi}^0+\dot{\xi}^1_{,i}\,.
\end{equation}
And the continuity, Euler equation and the evolution of dark matter density contrast are given accordingly
\begin{equation}
\dot{\delta} = -\theta + 3\dot{\Phi}
\end{equation}
\begin{equation}
\dot{\theta}+ {\cal{H}}= -\nabla^2\Psi
\end{equation}
\begin{equation}
\ddot{\delta} + {\cal{H}}\dot{\delta} - \frac{16\pi G a^{-1}}{3\xi^0(\tau)} \bar{\rho}\delta=0.
\end{equation}

\end{widetext}

\end{document}